\documentclass[11pt]{article}%
\usepackage{makeidx, siunitx, ragged2e}
\usepackage{booktabs,lscape,color, float}
\usepackage{subcaption} 
\usepackage{amsfonts,bm}
\usepackage{amsmath}
\usepackage{amssymb}
\usepackage{graphicx}
\usepackage{rotating}
\usepackage{multirow}%
\usepackage{geometry,setspace}
\usepackage{natbib}
\begin{document}
%%%%%%%%%%%%%%%%%%%%%%%%%%%%%%%%%%%%%%%%%%%%%%%%%%%%%%%%%%%%%%%%%%%%%%%%%%%%%%

\title{Equity Premium Puzzle or Faulty Economic Modelling?}
\author{
Abootaleb Shirvani\thanks{Texas Tech University, Department of Mathematics
\& Statistics, abootaleb.shirvani@ttu.edu.} \and Stoyan V. Stoyanov\thanks{Stony Brook University, College of Business stoyan.stoyanov@stonybrook.edu.}
  \and
Frank J. Fabozzi\thanks{EDHEC Business School, frank.fabozzi@edhec.edu.}
\and
Svetlozar T. Rachev\thanks{Texas Tech University, Department of Mathematics
	\& Statistics, zari.rachev@ttu.edu.}
}
\maketitle

\begin{abstract}
In this paper we revisit the equity premium puzzle reported in 1985 by Mehra and Prescott. We show that the large equity premium that they report can be explained by choosing a more appropriate distribution for the return data. We demonstrate that the high-risk aversion value observed by Mehra and Prescott may be attributable to the problem of fitting a proper distribution to the historical returns and partly caused by poorly fitting the  tail of the return distribution. We describe a new distribution that better fits the return distribution and when used to describe historical returns can explain the large equity risk premium and thereby explains the puzzle. 
\end{abstract}


%%%%%%%%%%%%%%%%%%%%%%%%%%%%%%%%%%%%%%%%%%%%%%%%%%%%%%%%%%%%%%%%%%%%%%%%%%%%%%
%\doublespacing
\newpage
%\part{Here we write the paper...}
%%%%%%%%%%%%%%%%%%%%%%%%%%%%%%%%%%%%%%%%%%%%%%%%%%%%%%%%%%%%%%%%%%%%%%%%%%%%%% 
\section{Introduction}\label{sec:intro}
\noindent An important measure in allocating funds among asset classes is the  risk premium of an asset class (i.e., the spread between the return on the asset class and the risk-free interest rate).  For this reason, there has been considerable research on the risk premium, particularly for equities. A study focusing on equities by \cite{Mehra:1985} found that for the US for the period 1889-1978 there was an excessively large equity risk premium relative to what would be expected if investors’ behavior toward risk followed what was assumed by proponents of rational finance. That is, investors were much more risk-averse than traditional finance models assumed. This finding was referred to by Mehra and Prescott as the ``equity premium puzzle”. 
The proponents of behavioral finance used the equity premium puzzle as an example of the limitations of rational finance. \cite{Benartzi:1995}, for example, suggested that “narrow framing” leads investors to overestimate equity risk and proposed an alternative to the standard investor preferences approach in Mehra and Prescott. They proposed the so-called myopic loss aversion model which is based on prospect theory, a theory based on experimental studies of human decisions
under risk rather than relying on the assumption of purely rational market participants.

\cite{Nada:2013} argues that the equity premium puzzle is attributable to the lack of consistency between the theoretical models employed and their calibrations on empirical data and a high level of relative risk aversion needed for both theory and empirics to coincide. Empirically, there is a concern about which theoretical models should be used to explain the large equity premium and the low risk-free rate.   
\cite{Mehra:2003} assumed that the growth rate of consumption and dividends are independent and identically log-normally distributed. They used the arithmetic mean in their analysis. In their model, the equity premium is the product of the coefficient of relative risk aversion (CRRA) and the variance of the growth rate of consumption. In this case, a high equity premium is impossible unless the CRRA is extremely high. This level of CRRA is consistent with a low risk-free rate and generates another puzzle; the risk-free rate puzzle. 

\cite{Kocherlakota:1996} discusses and assesses various theoretical attempts to explain the large equity premium and the low risk-free rate. He offers two explanations for the puzzle. The first is that there is a large differential between the cost of trading in the stock and bond markets. He argues that stock returns covary more with consumption growth than do Treasury bills.  Investors see stocks as a poorer hedge against consumption risk, and so stocks must earn a higher average return.
The second explanation according to \cite{Kocherlakota:1996} relates to the original parametric restrictions made by \cite{Mehra:1985}. He argues that the equity premium and risk-free rate puzzles are solely a product of the parametric restrictions imposed by Mehra and Prescott on the discount factor and the coefficient of relative risk aversion. He claims that individual investors have a CRRA larger than 10,\footnote{Two empirical studies more than 35 years apart support -- \cite{Friend:1975} and \cite{Chiappori:2011} – find support for the constancy of CRRA over time.  In contrast, using a GARCH-M model, \cite{Das:2010} find strong empirical evidence that CRRA varies over time. Following are estimates for the CRRA that have been reported by researchers: (1)  \cite{Friend:1975}, greater than 2, (2) \cite{French:1987}, 2.41, (3) \cite{Pindyck:1988}, range from 1.57 to 5.32, (4) \cite{Azar:2006}, 4.5, and  (5) \cite{Todter:2008}, 1.4 to 7.2.
} either with respect to their own consumption or with respect to per capital consumption.
Another avenue for resolving the puzzle has been to take into account rare events. \cite{Rietz:1988}, for example, appears to have been the first to do so. He   claimed to resolve the equity-premium puzzle by assuming low-probability disasters.  

In this paper,  we extend \cite{Mehra:2003} approach to accommodate rare events. We do so by assuming that the growth rate of consumption and dividends characterized by  heavy tails. More specifically, we assume that the distribution of the log-growth rate of consumption and dividends exhibits a normal inverse Gaussian (NIG) distribution.
We attempt to resolve the equity premium puzzle by fitting the NIG distribution to return data and critically evaluating the relative risk aversion estimate. We find that the CRRA estimate obtained from the NIG fitted model is significantly decreased compared to when the log-normal distribution is fitted, but it is not within the range that would be produced by the assumed investor attitude toward risk. 

From a technical point of view, our paper is an extension of the approach by  \cite{Lundtofte:2013}, who derived an exact expression for the equity premium under general distributions.  In their paper, \cite{Lundtofte:2013} worked with the NIG distribution and compared the CRRA to match the equity premium,
under normal and NIG distributions. They demonstrated that assuming a NIG
distribution helps mitigate the equity premium puzzle. The results in both \cite{Lundtofte:2013} and this paper offer empirical evidence in favor of assuming a NIG distribution. However in both papers, a high level of CRRA is needed for the historical equity premium to be consistent with theoretical models. This high level for CRRA indicates that the equity premium puzzle cannot be resolved by using the NIG distribution. 

The CRRA required to match the equity premium is significantly reduced by fitting the NIG distribution compared to the \cite{Mehra:2003} approach. This reduction in the estimate for the CRRA is due to using the NIG distribution and the resulting better fit in the presence of rare events. 
We believe that a distribution with the tails heavier than NIG can explain the puzzle. Thus, we reexamine the equity premium puzzle by defining a new distribution which we call the \textit{Normal compound inverse Gaussian} (NCIG) and evaluate the risk relative aversion. 
The estimated CRRA resulting from  fitting this distribution to the data is less than 10 and it is within the range that would be produced by the assumed investor attitude toward risk.

There are three sections that follow in this paper. In the next section, Section 2, we derive a formula for the log-equity premium by assuming a NIG distribution for the log-growth rate of consumption and dividends. In Section 3 we describe our data set and fit both the NIG and log-normal distributions to the data. We demonstrate that the high CRRA produced by the models is partially caused by poorly fitting the tail of the return distribution. In Section 4 we describe a new distribution and show how the proposed distribution is flexible
enough to offer an explanation for the equity premium puzzle. Section 5 concludes the paper.

\section{Equity Premium Puzzle}
In this section, we extend the approach by \cite{Mehra:2003} in order to accommodate rare events by assuming that the growth rate of consumption and dividends has a heavy-tailed distribution. 
We allow for a broad spectrum of distributional tails so that the statistical analysis of the data determines the type of distribution for the time period being analyzed. What we will show is that the distribution of the growth rate of consumption and dividends is highly unlikely to be log-normal. A much more flexible class of distribution is needed when modeling growth rates. To this end, we suggest the NIG distribution introduced by \cite{Barndorff:1977}.  \\

Random variable $X$  has a NIG distribution, denoted $X\sim NIG\left( \mu,\alpha,\beta,\delta\right) $, $  \mu \in \mathbb{R},\,\, \alpha\in \mathbb{R},\,\beta \in \mathbb{R},\,\delta \in \mathbb{R},\, \alpha^2>\beta^2$,
if its density is given by

\begin{equation}
\label{NIG_dis}
f_{X} \left( x\right) = \frac{\alpha \delta K_1 \left( \alpha \sqrt{ \delta^2 + \left( x-\mu \right) ^2 }\right) } { \pi \sqrt{\delta^2+\left( x-\mu\right) ^2 }} \exp {\left(  \delta \sqrt{ \alpha^2-\beta^2 }+\beta \left( x-\mu \right)\right) } , \,\,x \in \mathbb{R}.
\end{equation}

Then, $X$ has mean $E\left( X\right) =\mu +\frac{\delta \beta}{ \sqrt{\left( \alpha^2 -\beta^2 \right) }}$, variance $Var\left( X\right) =\frac{ \delta \alpha^2 }{\sqrt{\left( \alpha^2 -\beta^2 \right) ^{3} }}$ , skewness $\gamma\left( X\right) =\frac{3\beta}{\alpha \sqrt[4]{\delta^2 \left( \alpha^2-\beta^2\right) }}  $and excess kurtosis $\kappa\left( X\right) =\frac{3\left( 1+\frac{ 4\beta^2} {\alpha^2 }\right) }{\delta\sqrt{ \alpha^2-\beta^2 } }$. The characteristic function $\varphi_{X} \left( t\right) =E\left( e^{itX}\right), \, t\in \mathbb{R} $, is given 
by

\begin{equation}
\label{NIG_Char}
\varphi_{X}=\exp\left( {i\mu t +\delta\left( \sqrt{\alpha^2-\beta^2}-\sqrt{\alpha^2 - \left( \beta+it\right)^2 }\right) }\right) 
\end{equation}

Because the normal distribution, $N(\mu ,\sigma ^{2})$, is a special case of NIG by setting $\beta =0$, $\delta =\sigma ^{2}\alpha$, and letting $\alpha \rightarrow \infty$ (limiting case), We shall now replace the log-normal \footnote{$X$ is log-normally distributed, denote $X \sim logN\left( \mu,\sigma^2\right)$,  if $logX$ is normally distributed, $ logX \sim N 	 \left(\mu,\sigma^2 \right)$, with mean $\mu$ and variance $\sigma^2$.} assumption in \cite{Mehra:1985} with log-NIG, and obtain the result. 
What is more important is that by using the log-NIG distribution the result will be flexible enough to give the statistical-distributional explanation for the equity premium puzzle.

Let us briefly sketch the Mehra-Prescott model. It assumes a frictionless economy with one representative investor $\ell$  seeking to optimize the expected utility $E_0 \left( \sum_{t=0}^{\infty}b^t \mathbb{U}\left( C_t\right) \right) $,  where $\left( b \in \left(0,1 \right)\right) $ is the discount factor and $\mathbb{U}\left( c_t \right) $ is the utility from the consumption amount $c_t$ at time $\left( t=0,1,2,…\right) $. The utility function is given by  $\mathbb{U}\left( c\right) =\mathbb{U}^{\left( a\right)} \left( c\right) =\frac{c^{1-a}-1}{1-a}$, where $a>0$ is the CRRA. The agent invests in the asset at time $t$ giving $p_t$   units of consumption and sells the asset at $t+1$, receiving $p_{t+1}+y_{t+1}$, where $p_{t+1}$ is the asset price at $t+1$,  and $y_{t+1}$ is the dividend at $t+1$. In the Mehra-Prescott model, The agent's return on investment in  $(t,t+1]$ is given by

\begin{equation}
\label{Gross_Return}
R^{e}\left( t+1\right) =\frac{p_{t+1 }+\, y_{t+1 }}{p_t}=R^{f}\left( t+1\right) -
\frac{cov_t\left( \frac{\partial\mathbb{U}^{a}\left( c_{t+1 }\right) }{\partial c},R^{e}\left(t+1 \right) \right)} {E_t \left( \frac{\partial \mathbb{U}^{a}\left(c_{t+1} \right) }{\partial c}\right) }
\end{equation}
where $R^{f} \left( t+1\right)$  is the risk-free rate at $t+1$. Mehra and Prescott defined consumption growth in $(t,t+1]$ as $x_{t+1}=c_{t+1}/c_{t}$. Thus, The agent's return and the risk-free rate are

\begin{equation}
\label{Gross_Return2}
R^{e}\left(  t+1\right) =\frac{E_t\left( x_{t+1}\right) }{bE_t\left( x^{1-a}_{t+1}\right) },\,\,\,
\end{equation}
and 
\begin{equation}
\label{Gross_Risk}
R^{f}\left(  t+1\right) =\frac{1 }{bE_t\left( x^{-a}_{t+1}\right) }.
\end{equation}

\cite{Mehra:2003} make the following additional assumptions:

\begin{itemize}
	\item the growth rate of consumption $x_{t+1}=c_{t+1}/c_{t}$, $t=1,2,…, $  are independent and identically distributed (i.i.d) with $x_t \sim \ln \mathbb{N}\left( \mu^{x},\left( \sigma^{x}\right) ^{2}\right) $ or $\ln{x_t}\sim \mathbb{N}\left(\mu^{x},\left( \sigma^{x}\right) ^{2} \right) $. 
	\item the growth rate of dividends $z_{t+1}=y_{t+1}/y_{t}$, $t=1,2,…, $  are i.i.d.
	\item	$\left( x_t,z_t\right) $ are jointly log-normally distributed.
\end{itemize}

By imposing the equilibrium condition that x = z, a consequence is the restriction that the return on equity is perfectly correlated with the growth rate of consumption. Moreover, it leads to the following expression for the equity premium

\begin{equation}
\label{Equity_Premium}
\ln {E\left( R^{e}\left(  t+1\right)\right) }-\ln {R^{f}\left(  t+1\right)} = a \left( \sigma^{x}\right)^{2}
\end{equation}
Thus, the equity premium is equal to  CRRA times the variance of consumption growth. Testing their model on U.S. data for the period of 1889 to 1978, Mehra and Prescott found that $a$ is large and a high equity premium is impossible. As explained earlier, there is empirical evidence from several studies that the CRRA $a$ is less than 10.

We now assume that $\ln{x_t}\sim NIG\left( \mu,\alpha,\beta,\delta\right) $. From equation \eqref{Gross_Return2} it follows that 
\begin{equation}
\label{E_Return}
E\left( R^{e}\left(t+1 \right) \right) =\frac{\exp\left(\mu+\delta\left( \sqrt{\alpha^2-\beta^2}-\sqrt{\alpha^2-\left( \beta+1\right) ^2}\right)  \right) }{b\,\exp\left(\mu\left(1-a \right) +\delta\left( \sqrt{\alpha^2-\beta^2}-\sqrt{\alpha^2-\left( \beta+1-a\right) ^2}\right)  \right)},
\end{equation}
and 
\begin{equation}
\label{E_Risk}
R^{f}\left(t+1 \right) =\frac{1}{b\,\exp\left(\mu\left(-a \right) +\delta\left( \sqrt{\alpha^2-\beta^2}-\sqrt{\alpha^2-\left( \beta-a\right) ^2}\right)  \right)}.
\end{equation}
Thus, we have the following extension of the Mehra-Prescott equity premium:

\begin{equation}
\label{Equity_NIG}
\begin{array}{ccc}
\ln {E\left( R^{e}\left(t+1 \right) \right)}-\ln{R^{f}\left(t+1 \right)}= \\
\delta \left( \sqrt{\alpha^2-\beta^2}-\sqrt{\alpha^2-\left( \beta-a\right) ^2}-\sqrt{\alpha^2-\left( \beta+1\right) ^2}+\sqrt{\alpha^2-\left( \beta+1-a\right) ^2}\right),
\end{array}
\end{equation}
when  $\beta=0$ and $\delta=\alpha\left( \sigma^{x}\right) ^{2}$, then $\alpha\uparrow\infty$,

\begin{equation}
\label{Equity_Sim}
\begin{array}{ccc}
\ln {E\left( R^{e}\left(t+1 \right) \right)}-\ln{R^{f}\left(t+1 \right)}=\\ \alpha\left( \sigma^{x}\right) ^{2} \left( \alpha-\sqrt{\alpha^2- a^2}-\sqrt{\alpha^2-1}+\sqrt{\alpha^2-\left( 1-a\right) ^2}\right) \longrightarrow \alpha\left( \sigma^{x}\right) ^{2}.
\end{array}
\end{equation}
That is, we obtain Mehra-Prescott’s equity premium given by equation \eqref{Equity_Premium} as a limiting case of equation \eqref{Equity_NIG}. 

In order to compare equations \eqref{Equity_Premium} and \eqref{Equity_NIG} we standardize equation \eqref{Equity_Premium} with $\left( \sigma^{x}\right) ^{2}=1$  and equation  \eqref{Equity_NIG}  with  $\beta=0$, and $\delta=1$. Then $E\left( \ln{x_{t+1 }} \right)=\mu$, variance $var\left( \ln{x_{t+1}}\right)=1 $, skewness $\gamma\left( \ln{x_{t+1}}\right)=0 $, and excess kurtosis $\kappa \left( \ln{x_{t+1}}\right)=\frac{3}{|\alpha | } $. Consider then the ratio of the right-hand sides of equations  \eqref{Equity_Premium} and  \eqref{Equity_NIG}:
\begin{equation}
\label{Ratio}
\begin{aligned}
R\left( a,\alpha\right) =\frac{ \alpha \left( \alpha-\sqrt{\alpha^2- a^2}-\sqrt{\alpha^2-1}+\sqrt{\alpha^2-\left( 1-a\right) ^2}\right) }{a}.
\end{aligned}
\end{equation}

Here we note that there are numerous distributions with a heavy-tailed property. However to have an explicit formula for the equity premium given by  \eqref{Equity_Premium}, we should consider a distribution with a moment-generating function (MGF) having an exponential form. The reason for using the NIG distribution is that it is a fat-tail distribution, and its MGF has an exponential form.

\section{Data}
The data used in this study are the monthly  historical adjusted-closing price for the S\&P 500 and the dividend of the S\&P 500 from 1900-2018 collected from the Bloomberg Financial Markets, and the consumption of non-durable goods and services data from 1960 to 2018 obtained from the Federal Reserve Economic Data website.  The 10-year Treasury bill rate is used as the risk-free rate and obtained from the economic research division of the Federal Reserve Bank of St. Louis from 1900-2018. The average annual real return from 1900 to 2018 is 6.81\% for the S\&P 500 and 0.987\% for the risk-free asset.  Therefore, the mean U.S. equity premium is 5.894\%. 

\section{Model Validation and Results}
\subsection {The NIG and normal model with risk aversion coefficient estimation}
Here we discuss fitting the normal and NIG distributions to historical data and evaluate which distribution best fits the data set. The maximum-likelihood estimation (MLE) method is used to  estimate the parameters for both distributions. The estimated parameters and log-likelihood values  are reported in Table \ref{tab:parameter estimation}.\footnote{ The values were estimated using the R-Package GeneralizedHyperbolic. See \cite{Scott:2015a}.}

\begin{table}[h]
	\centering
	\caption{Estimation of the NCIG model parameters for the growth rate of consumption}
	\begin{tabular}{@{}llllll@{}}
		\toprule
		Distribution & \multicolumn{4}{c}{Parameter}                                                & Log-Likelihood \\ \midrule
		normal       & \multicolumn{2}{l}{$\mu^{x}=0.0014508893$} & \multicolumn{2}{l}{$\sigma^{x}=0.0128671292$} & 4192.89       \\
		NIG          & $\mu=0.002351$     & $\delta=0.006590$    &$ \alpha=38.437308$    & $\beta=-5.194172$    & 4444.44        \\ \bottomrule
	\end{tabular}
	\label{tab:parameter estimation}%
\end{table}

\begin{figure}[b!]
	\centering
	\begin{tabular}{lrlr}
		(a)&	
		\includegraphics[scale = 0.25]{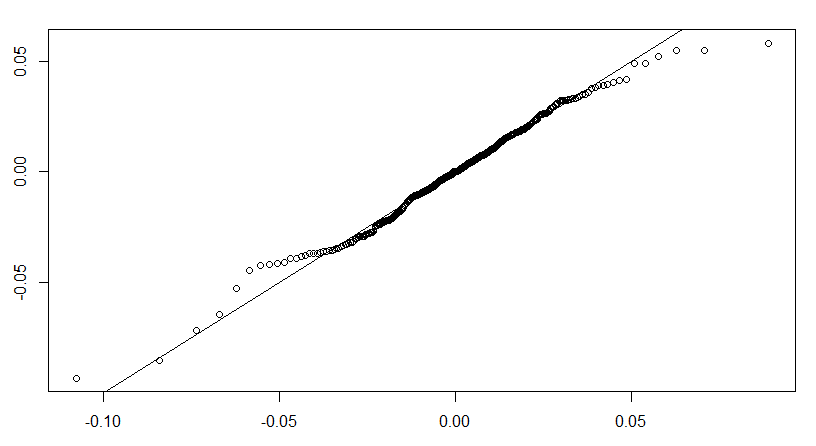} &
		(b)&
		\includegraphics[scale = 0.25]{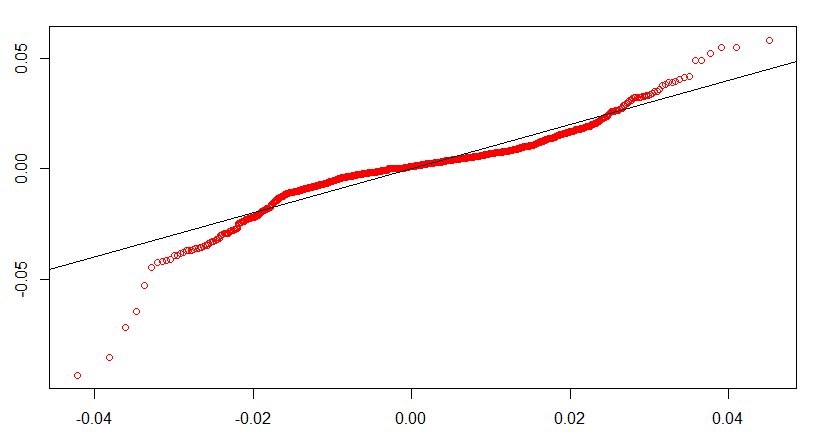}\\
		(c)&
		\includegraphics[scale = 0.25]{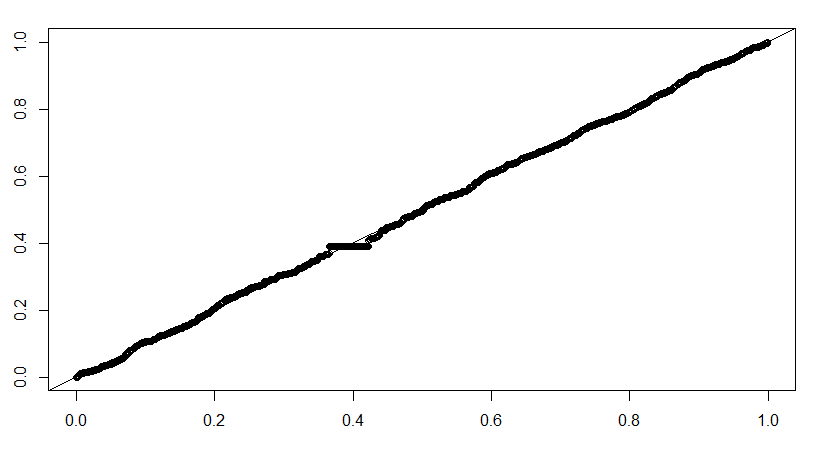}&
		(d)&
		\includegraphics[scale = 0.25]{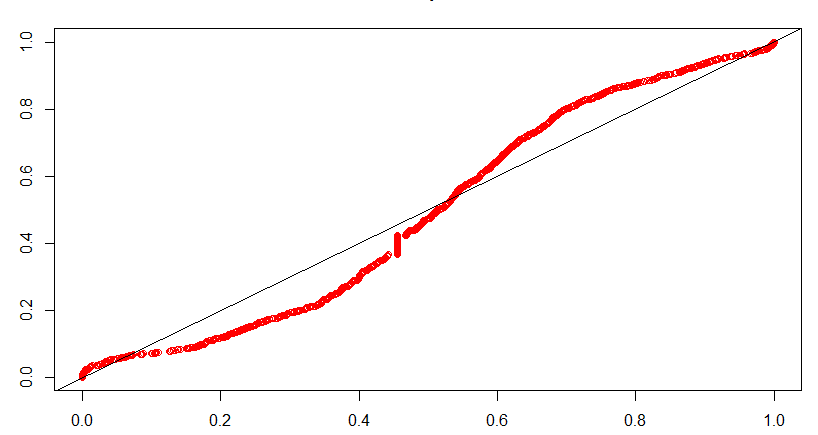}
	\end{tabular}
	\caption{(a) Q-Q plots of fitted NIG distribution, 
		(b) Q-Q plots of fitted normal distribution,  (c) P-P plots of fitted NIG distribution, and  
		(d) P-P plots of fitted normal distributions  }
	\label{QQ_plot}
\end{figure}

The higher log-likelihood value for the NIG distribution indicates that this distribution is a better fit than  the normal distribution. Figure $\ref{QQ_plot}$ shows the Q-Q and P-P plots for the NIG and normal distributions, respectively. At first glance the linearity of the NIG P-P plot appears to validate our choice of NIG as the theoretical distribution. The quantiles for the data sets for the normal distribution (P-P plot) do not nearly match the line given by the Q-Q plot for the NIG distribution. This means that there is a greater concentration of data beyond the left and right tails of a normal distribution. It shows that the data set are fat in the tails and therefore one needs to fit a distribution to the data set with a tail heavier than the normal distribution. Thus assuming a fat tail distribution for the data set would be more suitable. Accordingly, in conjunction to the data set with the normal distribution, the NIG fit appears to be a proper distribution for fitting the data set. Therefore, we use the NIG model to look at the equity puzzle.

\begin{figure}[t!]
	\centering
	\includegraphics[width=.75\textwidth]{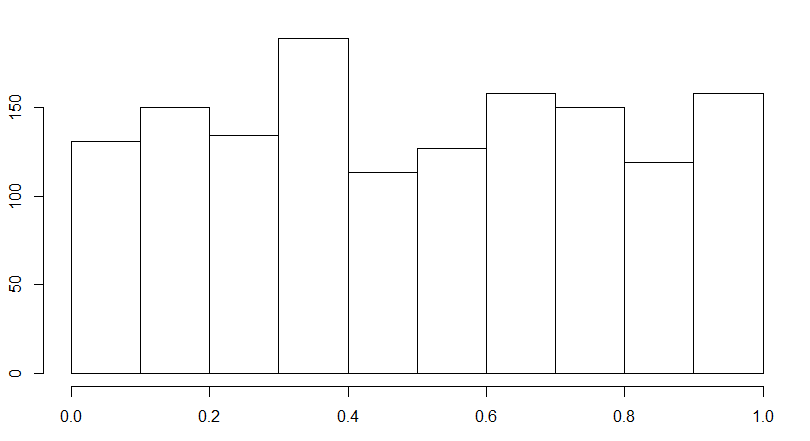}
	\caption{Probability integral transformation histogram. }
	\label{histo_uni}
\end{figure}

In addition, we evaluated the fitted distribution by the probability integral transformation test. We perform probability integral transformation to map our data set to interval $\left(0,1 \right)$ through the cumulative   distribution function (CDF); that is, $Y=F_X\left( x\right) $ which is uniform density \citep[see][]{Diebold:1998}. The uniformity of the probability integral transform is evaluated using the histogram plot and goodness of fits such as the Kolmogorov-Smirnov, Neyman, and Frosini tests. The histogram diagram for the CDF (see Figure  $\ref{histo_uni}$) shows that the probability integral transform comes from the uniform distribution. Kolmogorov-Smirnov, Neyman, and Frosini tests support the uniformity of probability integral transformation at the 5\% confidence level since the p-value of all three tests is greater than  $0.05$ (Kolmogorov-Smirnov, 0.58; Neyman, 0.89, and; Frosini, 0.76).

In Section 3, we reported that the mean U.S. equity premium is 5.894\% based on the data set going back to 1900. The mean annual real return from 1900 to 2018 is 6.81\% for the S\&P 500 and 0.987\% for the risk-free asset. Having the mean return equity premium, risk-free rate, and the NIG parameters by calibrating the equation \eqref {Equity_NIG} the estimate for the CRRA is $33.5$. When the normal distribution is fitted, the estimate for the CRRA is 2582.6. This reduction in the estimate for the CRRA is due to using the NIG distribution and the resulting better fit to rare events. \cite{Mehra:2003} quoted that the CRRA is a small number, and the most of studies indicate that is bounded from above by 10. Thus, our finding for the CRRA is not within the range that would be produced by the assumed investor attitude toward risk. 

If we set CRRA equal to 10 in equations (7) and (8), the mean return for the S\&P 500 is 1.3439 and for the risk-free asset  is 1.3417. This implies an equity risk premium 0.2223\%, far lower than the historical equity premium (5.894\%). In the both fitted models, a high level of CRRA is needed for the historical equity premium to be consistent  with theoretical models. This finding indicates that the equity premium puzzle cannot be resolved  by using the NIG distribution. Therefore, there is a problem in fitting the NIG distribution that is reflected by the high value for CRRA.

\subsection {The normal compound inverse Gaussian model with  relative risk aversion coefficient}

Although the NIG P-P plot in Figure~\ref{QQ_plot} shows a well-fitted distribution to the data, the Q-Q plot indicate that both models fitted poorly in areas of low density or in the tails. Thus, there is a problem in fitting the model and this problem is at least partly caused by the tail of the distributions. Therefore, in our applications, a well-fitted model to rare events is the main concern and therefore a fat-tail distribution is needed to capture the extreme events.

In this subsection, we introduce a new type L\'{e}vy process relative to NIG which we call the \textit{normal compound inverse Gaussian} (NCIG) and use that distribution to  estimate relative risk aversion. The NCIG is a mixture of the normal and doubly compound of the inverse Gaussian (IG) distribution.\footnote{For a further discussion of the use of NCIG distribution and double subordinator models, see \cite{Shirvani:2019}.  } It is a  heavy-tailed distribution with tails that are heavier compared to the NIG distribution. We will show that the MGF of the NCIG distribution has an exponential form. To have an explicit formula for the equity premium given by equations \eqref{Gross_Return2} and \eqref{Gross_Risk}, the main driver of our consideration is a distribution with MGF with an exponential form. It seems that the NCIG distribution is an efficient distribution in our work because  it is a heavy-tail distribution and its MGF has an exponential form. 

To define the NCIG distribution, we first describe some features  of the IG processes. Random process $T(t), \,t\geq0$  has IG distribution \citep[see][]{Chhikara:1989}, denoted by $X\sim IG\left( \lambda_T,\mu_T\right) $, for some shape parameter $\lambda_T>0$ and mean parameter $\mu_T>0$ if its density is given by

\begin{equation}
\label{IG_dis}
f_{T_t} \left( x\right) = \sqrt { \frac{\lambda_T}{2\pi x^3} } \, \exp{\left( -\frac{\lambda_T\,\left( x-\mu_T\right)^3 }{2\mu_{T}^2\,x}\right)  }.
\end{equation}

\noindent \textit{Definition 1: Doubly Subordinated IG Process:}
Let $T(t)$ and $U(t)$, be independent IG L\'{e}vy subordinators,\footnote{A L\textrm{\'{e}}vy subordinator is a L\textrm{\'{e}}vy process with increasing sample path \citep[see][]{Sato:2002}.} $ T(1)\sim\,IG\left( \lambda_T,\mu_T\right)$, $U(1)\sim\,IG\left( \lambda_U,\mu_U\right) $, then the  compound subordinator $V(t)=T\left( U(t)\right)$ has density function %$f_{V(1)}(x)$%, 
given by

\begin{equation}
\label{Doubly_Subordinated}
f_{V(1)}(x)=\frac{1}{2\pi}\sqrt{\frac{\lambda_T\lambda_U}{x^3}}\int_{0}^{\infty}u^{-\frac{3}{2}}\exp\left(-\frac{\lambda_T\left(x-\mu_T u \right)^2 }{2\mu_{T}^2 x}-\frac{\lambda_U\left(u-\mu_{U} \right)^2 }{2\mu_{U}^2 u} \right)du ,
\end{equation}
where $x>0$. The MGF of $V(1)$ is
\begin{equation}
\label{MGF_IG}
M_{V(1)}(v)=\,\exp\left( \frac{\lambda_U}{\mu_{U}} \left(1-\sqrt{1-2\,\frac{\mu_{U}^2\,\lambda_T}{\lambda_U\,\mu_{T}} \left(  1-\sqrt{1-\frac{2\mu_{T}^2}{\lambda_T}v}\right)}\right)\right),
\end{equation}
where $v\in\,\left( 0,\frac{\lambda_T}{2\mu_{T}^2}\left[ 1-\left( \frac{\lambda_T\,\mu_{T}}{2\mu_{U}^2\,\lambda_U}\right)^2 \right] \right)$.	
If $T(t)$ is a L\'{e}vy process with L\'{e}vy exponent, $\psi_{T}(u)\,=\,-\ln\,E\left[\exp\left(iu\,T(1) \right)\right]$ $u\in \mathbb{R}$, and $U(t)$, independent of $T(t)$, is a L\'{e}vy subordinator with Laplace exponent $\phi_{U}(s)\,=\,-\ln E\left[\exp\left( -s\,U(1)\right)\right],\,\,s>0$ then the subordinator process $Y(t)=T\left(U(t) \right)$ is again a L\'{e}vy process with L\'{e}vy exponent and probability transition  given \footnote{See Chapter 6 of \cite{Sato:2002}.}

\begin{equation}
\label{levy_exponent}
\begin{array}{cc}
\psi_{Y}(u)\,=\,\phi_{T}\left( \psi_{U}(u)\right),
\end{array}
\end{equation}
\begin{equation}
\label{probability_transition}
P_{Y}(t,A)=\,\int_{0}^{\infty} P_{T}(t,A)\,P_{U}(u,dt),
\end{equation}
respectively. 
In particular, if $B_t$ is a Brownian motion, denoted by $B^{	\nu,\sigma^2}$, with L\'{e}vy exponent,   $\psi_{B}(u)\,=\,-\ln\,E\left[\exp\left(iu\,B(1)    \right)  \right]=\,-iu	\nu+\frac{\sigma^2}{2}u^2$, and $T(t)$, independent of $B(t)$, is an IG-subordinator with Laplace exponent given
\begin{equation}
\label{laplace_exponent_IG}
\phi_{T(1)}(s)=-\ln E\left[ \exp\left(-s\,T(1) \right) \right]=\,-\frac{\lambda_T}{\mu_{T}}\left(1-\sqrt{1+\frac{2\mu_{T}^2 \,s}{\lambda_T}} \right) , 
\end{equation}
then $Y(t)=B(T(t))$ is NIG process with L\'{e}vy exponent 
\begin{equation}
\label{laplace_NIG}
\psi_{Y}(u)=\phi_{T}\left(\psi_{B}(u) \right) =\,-\frac{\lambda_T}{\mu_{T}} \left(1-\sqrt{1+\frac{2\mu_{T}^2 }{\lambda_T}\left(-iu	\nu+\frac{\sigma^2}{2}u^2 \right) } \right) .
\end{equation}

\noindent \textit{Corollary:} Let $T(t)$ be an IG subordinator with L\'{e}vy exponent 
\begin{equation}
\label{levy_exponent_IG}
\psi_{T(1)}(u)=-\ln E\left[ \exp\left(iu\,T(1) \right) \right]=\,-\frac{\lambda_T}{\mu_{T}}\left(1-\sqrt{1-\frac{2\mu_{T}^2 \,iu}{\lambda_T}} \right), 
\end{equation}
and $U(t)$, independent of $T(t)$, be an IG process with Laplace exponent given
\begin{equation}
\label{IG_Laplace_exponent}
\phi_{U(1)}(s)=-\ln E\left[ \exp\left(-s\,U(1) \right) \right]=\,-\frac{\lambda_U}{\mu_{U}}\left(1-\sqrt{1+\frac{2\mu_{U}^2 \,s}{\lambda_U}} \right), 
\end{equation}
then $V(t)=T\left(U(t) \right)$ is a subordinator with L\'{e}vy exponent given 
\begin{equation}
\label{levy_dobly_IG}
\psi_{V(1)}(u)=\phi_{U(1)}\left(\psi_{T}(u) \right) 
=-\frac{\lambda_U}{\mu_{U}} \left(1-\sqrt{1-2\,\frac{\mu_{U}^2\,\lambda_T}{\lambda_U\,\mu_{T}} \left(  1-\sqrt{1-\frac{2\mu_{T}^2}{\lambda_T}iu}\right)}\right).
\end{equation}
Proof: Using \eqref{levy_exponent}, and substituting \eqref{levy_exponent_IG} in \eqref{IG_Laplace_exponent}. 

A special case of the NCIG distribution is when $\lambda_T=\lambda_U=\lambda$ and $\mu_{T}=\mu_{U}=\mu$. Suppose $\lambda_T=\lambda_U=\lambda$ and $\mu_{T}=\mu_{U}=\mu$, then the Laplace exponent of the doubly IG subordinator, $V(t)=T(U(t))$  is  
\begin{equation}
\label{Laplace_exponent_IG}
\phi_{V(1)}(s)=-\ln\,E\left[\exp\left(-s\,V(1)\right)\right]=-\frac{\lambda}{\mu}\left(1-\sqrt{1-2\mu\left(  1-\sqrt{1+\frac{2\mu^2}{\lambda}s}\right)}\right),
\end{equation}
and for each $v\in\,\left( 0\,,\frac{\lambda}{2\mu^2}\left[ 1- \frac{1}{4\mu^4} \right] \right)$,
the MGF is 
\begin{equation}
\label{Moment_dublly_IG}
M_{V(1)}(v)=\exp\left( \frac{\lambda}{\mu} \left(1-\sqrt{1-2\,\mu \left(  1-\sqrt{1-\frac{2\mu^2}{\lambda}v}\right)}\right)\right).
\end{equation}

\noindent \textit{Definition 2. Normal compound inverse Gaussian:}
Let $V(t)=T(U(t))$ be a doubly subordinator IG process with MGF and Laplace exponent given by equations \eqref{Moment_dublly_IG} and \eqref{Laplace_exponent_IG} respectively, and $B(t)$ is a Brownian motion L\'{e}vy process, denoted by $B^{	\nu,\sigma^2}$, then the L\'{e}vy process $Z(t)=B^{	\nu,\sigma^2}\left(V(t) \right)$  is a NCIG L\'{e}vy process, denoted $Z(t)\sim NCIG(\mu,\lambda,	\nu,\sigma^2)$, with L\'{e}vy exponent given by
\begin{equation}
\begin{array}{ccc}
\label{levy_exponent_NCIG}
\psi_{Z}(u)=\phi_{V}\left(\psi_{B}(u) \right) =-\frac{\lambda}{\mu} \left(1-\sqrt{1-2\,\mu \left(  1-\sqrt{1+\frac{2\mu^2}{\lambda}\psi_{B^{	\nu,\sigma^2}}(u)}\right)}\right)\\
=- \frac{\lambda}{\mu} \left(1-\sqrt{1-2\,\mu \left(  1-\sqrt{1-\frac{2\mu^2}{\lambda}\left(iu	\nu-\frac{\sigma^2}{2}u^2 \right) }\right)}\right).
\end{array}
\end{equation}
The characteristic function (ch.f.) of $Z(1)$ is   
\begin{equation}
\label{Char_NCIG}
\varphi_{Z(1)}(u)= %E\left[ \exp\left(iu\,Z(1) \right) \right]=
%\exp\left( -\psi_{Z}(u)\right) =
\exp\left( \frac{\lambda}{\mu} \left(1-\sqrt{1-2\mu \left(1-\sqrt{1-\frac{2\mu^2}{\lambda}\left(iu	\nu-\frac{\sigma^2}{2}u^2\right)}\right)}\right)\right) ,
\end{equation}
and the MGF is
\begin{equation}
\label{MGF_NCIG}
M_{Z(1)}(s)= %E\left[ \exp\left(s\,Z(1) \right) \right]=
\exp\left( \frac{\lambda}{\mu} \left(1-\sqrt{1-2\,\mu \left(  1-\sqrt{1-\frac{2\mu^2}{\lambda}\left( s	\nu+\frac{\sigma^2}{2}s^2\right) }\right)}\right)\right),
\end{equation}
where $s>0$. Furthermore, $2\mu\left(1-\sqrt{1-\frac{2\mu^2}{\lambda}\left(s\nu+\frac{\sigma^2}{2}s^2\right)}\right)<1$, i.e. $\frac{\sigma^2}{2}s^2+s	\nu-\frac{\lambda}{2\mu^3}\left( 1-\frac{1}{4\mu}\right) <0$.

Now, we assume the log-growth rate of dividend, by imposing the equilibrium condition, has NCIG distribution, $\ln x_t\sim NCIG(\mu,\lambda,a,\sigma^2)$. Then, from equation \eqref{Gross_Return2} it follows that 

\begin{equation}
\label{Return_NCIG}
\begin{array}{ll}
E\left(R^{e}(t+1) \right)  = 
\frac{\exp\left( \frac{\lambda}{\mu}\left( 1-\sqrt{1-\,2\mu\left( 1-\sqrt{1-\frac{2\mu^2}{\lambda}\left(	\nu+\frac{\sigma^2}{2} \right)  }\right) }\right)\right) } 
{b\exp\left( \frac{\lambda}{\mu}\left( 1-\sqrt{1-\,2\mu\left( 1-\sqrt{1-\frac{2\mu^2}{\lambda}\left((1-a)	\nu+\frac{\sigma^2}{2}(1-a)^2 \right)  }\right) }\right)\right)},
\end{array}
\end{equation}

where $a$ is the CRRA. Then similarly, the gross return on the risk-free asset from the equation \eqref{Gross_Risk} is  
\begin{equation}
\label{Riskless_NCIG}
R^{f}_{t+1}=\,\frac{1}{b\exp\left( \frac{\lambda}{\mu}\left( 1-\sqrt{1-\,2\mu\left( 1-\sqrt{1-\frac{2\mu^2}{\lambda}\left(-a 	\nu+\frac{\sigma^2\,a^2}{2} \right)  }\right) }\right)\right)} .
\end{equation}
Thus, we have the following extension of the Mehra-Prescott equity premium:
\begin{equation}
\label{equity_premium_NCIG}
\begin{array}{ll}
\ln E\left(R^{e}_{t+1} \right) -\ln\left(R^{f}_{t+1} \right) = &  \, \frac{\lambda}{\mu}\left( 1+\,A_1-\,A_2-A_3\right)  
\end{array}
\end{equation}
where, \\$A_1=\sqrt{1-\,2\mu\left( 1-\sqrt{1-\frac{2\mu^2}{\lambda}\left((1-a)	\nu+\frac{\sigma^2}{2}(1-a)^2 \right)  }\right) } $,\\
\noindent $A_2=\sqrt{1-2\mu\left(1-\sqrt{1-\frac{2\mu^2}{\lambda}\left(-a\nu+\frac{\sigma^2}{2}a^2 \right)} \right)}$,\\
\noindent and $A_3=\sqrt{1-\,2\mu\left( 1-\sqrt{1-\frac{2\mu^2}{\lambda}\left(	\nu+\frac{\sigma^2}{2}\right)  }\right) }$.

Instead of using the maximum likelihood method  to estimate the model parameters, we apply model fitting via the empirical characteristic function (ECF) \citep[see][]{Yu:2003}. The existence of a one-to-one correspondence between the
CDF and the ch.f. makes inference and estimation using the ECF method as efficient as the maximum likelihood method.\footnote{See \cite{Yu:2003}.} 
We obtain the initial values by using the method of moments estimation methodology.
For any initial value, we estimated the model parameters and consider the model as a good candidate to fit the data. We implemented the fast fourier transform  to calculate both the probability density function (pdf) and the corresponding likelihood values. The best model to fit and explain the observed data is chosen as the one with $(1)$ the largest likelihood value and $(2)$ the likelihood value greater than the NIG likelihood value.
The estimated value for the model parameters are reported in Table \ref{tab:NCIG}. 

\begin{table}[h]
	\centering
	\caption{Estimation of the NCIG model parameters}
	\begin{tabular}{@{}lllll@{}}
		\toprule
		Parameter  & \,\,\,\,\,\,\,$\lambda$\,\,\,  &\,\,\, $\mu$ \,\,\,  & \,\,\, $\nu$    \,\,\,& \,\,\,$\sigma^2$ \,\,\,\\ \midrule
		Estimation & 195.903 & 0.261 & 0.08 & 3.472                    \\ \bottomrule
	\end{tabular}
	\label{tab:NCIG}%
\end{table}

In our calibration from \eqref{equity_premium_NCIG}, the estimate for the CRRA is $8.9626$. This is markedly lower than what we obtained by fitting the normal distribution and NIG distribution.
In general, most economist believe that a risk-aversion coefficient above 10 reflects implausible behavior on the part of individuals. According to \cite{Nada:2013}, the evidence against a high relative CRRA is not that strong but this argument does not rescue the power utility model. That’s why a value of CRRA below 10 is acceptable. In our approach, by modifying the probability distribution to capture extreme events, the estimated CRRA declines from 33.55 to 8.9626 and the latter estimate is in the proposed range believed by the vast majority of economists.

\section{Conclusion}
In this study, we seek to demonstrate that the equity premium puzzle identified by  \cite{Mehra:1985} can be explained by fitting a more appropriate distribution for the growth rate of consumption and dividends. Prior explanations for the existence of the puzzle relied on arguments put forth by proponents of behavioral finance. We fitted the normal inverse Gaussian and log-normal distributions to the data and evaluated the relative risk aversion estimates. These estimates for the relative risk aversion for both models are markedly higher than what would be consistent with rational finance models. This is because estimates from both fitted distributions do not perform well due to their inability to deal with rare events. The estimate for the relative risk aversion is significantly lower when the normal inverse Gaussian distribution is fitted compared to when the log-normal distribution is fitted. The high estimated value for relative risk aversion reflects a problem in fitting the normal inverse Gaussian distribution. This problem is at least partly caused by the distribution tails that  are not enough heavy to fit in rare events.  

We argued that the abnormally large estimated value for relative risk aversion is reduced by fitting a heavy-tail distribution to the data. We introduced the normal compound inverse Gaussian distribution, a model with heavy tails, for modeling dividend returns. By fitting the normal compound inverse Gaussian distribution to the data, the relative risk aversion coefficient reduced to $8.9626$, a value that is within the range acceptable to economists. We conclude that using the new fitted distribution to historical return data can explain the large equity risk premium and thereby can explain the puzzle.

%%%%%%%%%%%%%%%%%%%%%%%%%%%%%%%%%%%%%%%%%%%%%%%%%%%%%%%%%%%%%%%%%%%%%%%%%%%

%%%%%%%%%%%%%%%%%%%%%%%%%%%%%%%%%%%%%%%%%%%%%%%%%%%%%%%%%%%%%%%%%%%%%%%%%%%
\end{document}